\documentclass[prd,preprintnumbers,twocolumn,eqsecnum,floatfix,nofootinbib,a4paper,superscriptaddress]{revtex4-1}
\usepackage{color}
\usepackage{calc}
\usepackage{amsmath,amssymb,graphicx}
\usepackage{amssymb,amsmath}
\usepackage{tensor}
\usepackage{bm}
\usepackage{float}
\usepackage{microtype}
\usepackage{booktabs}
\usepackage{times}
\usepackage[varg]{txfonts}
\usepackage[colorlinks, pdfborder={0 0 0}]{hyperref}
\usepackage{subfigure}
\definecolor{LinkColor}{rgb}{0.75, 0, 0}
\definecolor{CiteColor}{rgb}{0, 0.5, 0.5}
\definecolor{UrlColor}{rgb}{0, 0, 0.75}
\hypersetup{linkcolor=LinkColor}
\hypersetup{citecolor=CiteColor}
\hypersetup{urlcolor=UrlColor}
\allowdisplaybreaks
\usepackage[utf8]{inputenc}
\usepackage{ulem}
\usepackage{comment}
\usepackage{hyperref}
\usepackage[nolist]{acronym}
\usepackage{tikz}
\usepackage{textcomp}
\usepackage[export]{adjustbox}
\usepackage{url}

\newcommand{\zl}{z_{\ell}}

\newcommand{\mwdm}{m_{\wdm}}

\newcommand{\tobs}{T_{\mathrm{obs}}}
\newcommand{\wdm}{\textsc{wdm}}
\newcommand{\cdm}{\textsc{cdm}}

\usetikzlibrary{arrows,shapes,trees,decorations.pathreplacing}
\usepackage{aas_macros}
\usepackage{diagbox}

\allowdisplaybreaks

\begin{document}

\title{Probing the nature of dark matter using strongly lensed gravitational waves from binary black holes}

\author{Souvik Jana}
\affiliation{International Centre for Theoretical Science, Tata Institute of Fundamental Research, Bangalore 560089, India}

\author{Shasvath J.  Kapadia}
\affiliation{International Centre for Theoretical Science, Tata Institute of Fundamental Research, Bangalore 560089, India}
\affiliation{Inter-University Centre for Astronomy and Astrophysics, Post Bag 4, Ganeshkhind, Pune 411007, India}

\author{Tejaswi Venumadhav}
\affiliation{International Centre for Theoretical Science, Tata Institute of Fundamental Research, Bangalore 560089, India}
\affiliation{Department of Physics, University of California at Santa Barbara, Santa Barbara, CA 93106, USA}

\author{Surhud More}
\affiliation{Inter-University Centre for Astronomy and Astrophysics, Post Bag 4, Ganeshkhind, Pune 411007, India}

\author{Parameswaran Ajith}
\affiliation{International Centre for Theoretical Science, Tata Institute of Fundamental Research, Bangalore 560089, India}
\affiliation{ Canadian Institute for Advanced Research, CIFAR Azrieli Global Scholar, MaRS Centre, West Tower, 661 University Ave, Toronto, ON M5G 1M1, Canada}
	
\begin{abstract}
Next-generation ground-based gravitational-wave (GW) detectors are expected to detect millions of binary black hole mergers during their operation period. A small fraction ($\sim 0.1 - 1\%$) of them will be strongly lensed by intervening galaxies and clusters, producing multiple copies of the GW signals. The expected number of lensed events and the distribution of the time delay between lensed images will depend on the mass distribution of the lenses at different redshifts. Warm dark matter or fuzzy dark matter models predict lower abundances of small mass dark matter halos as compared to the standard cold dark matter. This will result in a reduction in the number of strongly lensed GW events, especially at small time delays. Using the  number of lensed events and the lensing time delay distribution, we can put a lower bound on the mass of the warm/fuzzy dark matter particle from a catalog of lensed GW events. The expected bounds from GW strong lensing from next-generation detectors are significantly better than the current constraints.
\end{abstract}

\maketitle
	
\paragraph*{Introduction:---} 

\begin{figure*}[tbh]
\includegraphics[width=0.68\columnwidth]{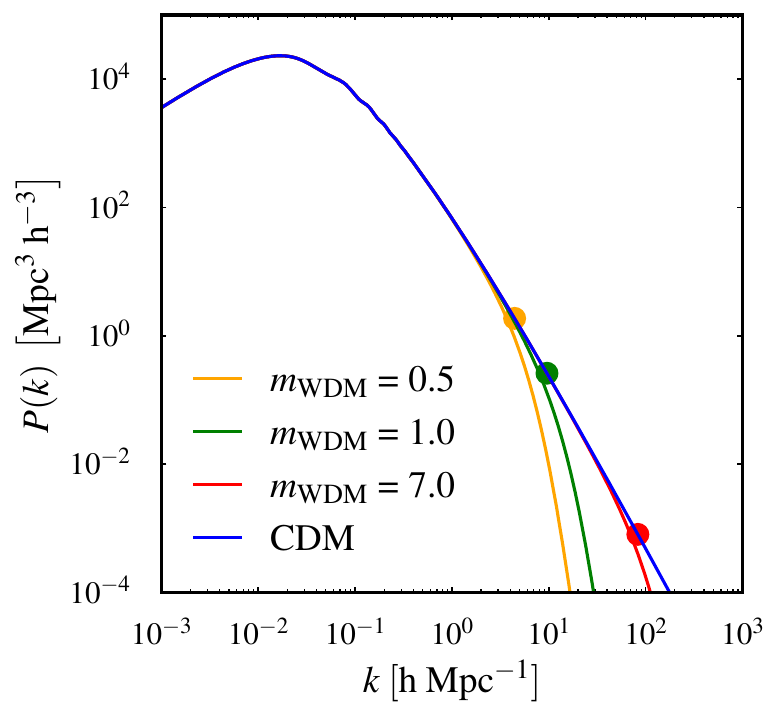}
\includegraphics[width=0.68\columnwidth]{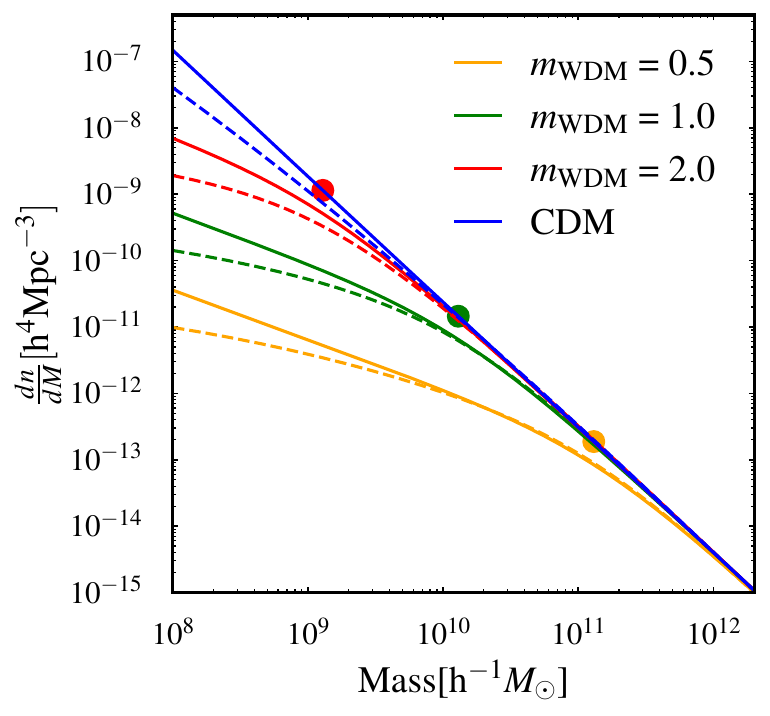}
\includegraphics[width=0.68\columnwidth]{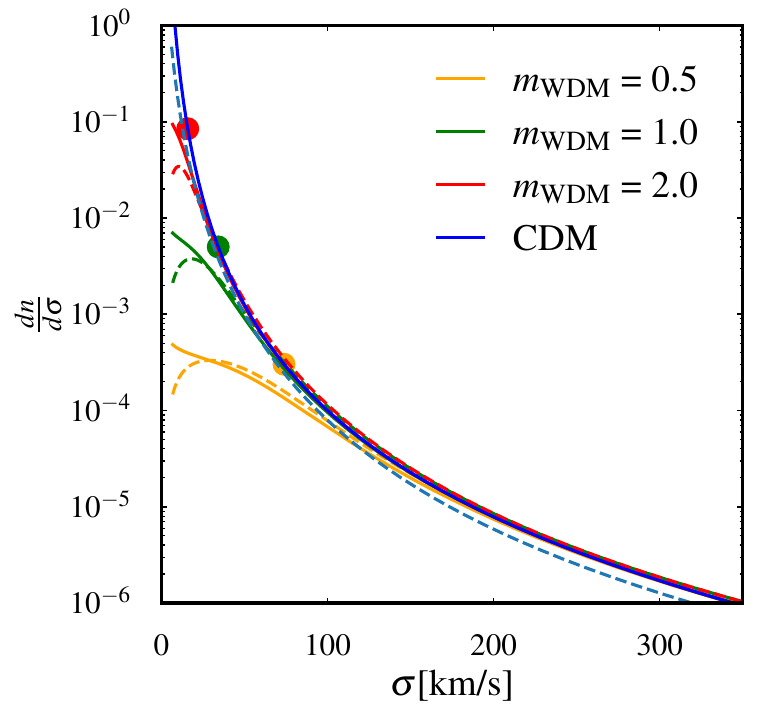}
\caption{\emph{Left panel:} The power spectrum of linear perturbations as predicted by the CDM model as well as the WDM model with different values of $m_\wdm$ (in keV). Half-mode scales for different $m_\wdm$ are shown by the filled circles. \emph{Middle panel:}  HMF for CDM and WDM at redshift $z=0$. Solid and dashed lines represent the Behroozi~\cite{behroozi2013} and Jenkins~\cite{jenkins2001} HMF models, respectively. Note the suppression in the number density of lower mass halos in the WDM model. Filled circles with different colours denote the half mode mass scale for different $m_\wdm$.  \emph{Right panel:} The distribution of the velocity dispersion of lenses produced by the CDM and WDM halos. Here also the solid and dashed lines represents the Behroozi and Jenkins HMF models. Reduction in number density is reflected as reduction of the low $\sigma$ halos. Filled circles denote the velocity dispersion of the corresponding half mode mass.}
\label{fig:hmf-sigma}
\end{figure*}

A variety of astronomical observations have firmly established that $\sim 25\%$ of the mass energy in the universe is in the form of some non-baryonic dark matter (DM)~\cite{Bertone_2018}. Particle physicists and cosmologists have come up with several candidates for DM, spanning a very wide mass range. The list of candidates ranges from extremely light elementary particles~\cite{Bertone_2018,ARBEY2021103865} to supermassive primordial black holes~\cite{Carr_2022}. 

DM candidates can be classified according to their velocity dispersion, which defines a free streaming length scale. Below this length scale all the cosmological density perturbations are wiped out, so no structure can form in the universe below this length scale. \emph{Cold} DM (CDM), such as the weakly interacting massive particles~\cite{Griest_2000}, axions \cite{Feng_2010} and  primordial black holes \cite{Green_2021}, has small free streaming length scales and does not affect the cosmological structure formation. On the other hand, \emph{hot} DM such as neutrinos is highly relativistic. Free streaming of such relativistic particles would erase perturbations in the matter density even on the scale of galaxy clusters ($\sim 10^{15}\mathrm{M}_{\odot}$). The very existence of such large scale structures has ruled out hot DM~\cite{White_1983}. In between there exists another class, called \emph{warm} DM (WDM), such as gravitino~\cite{Kawasaki_1997} and sterile neutrino~\cite{BOYARSKY20191}. They are non-relativistic but still have  non-negligible velocity dispersion. They have shorter free streaming length than regular neutrinos, and can erase structure on galaxy scales. Thus the existence of galaxies can put some (rather weak) constraints on the mass of the WDM particle.

In the past decades, the cosmological constant dominated CDM model known as $\Lambda$CDM, has emerged as the standard model of cosmology~\cite{Komatsu_2011, survey_2dFGRS_2005, SDSS_2006}. However, despite decades of effort, neither direct laboratory experiments or indirect astronomical observations have been able to detect any CDM candidates so far. In addition, though $\Lambda$CDM predictions match with observed large scale structure, sub-galactic observations might be in conflict with the predictions of this model. One is the apparent under-abundance of satellites in the Milky Way, as compared to the earlier CDM simulations, called the ``missing satellite problem''~\cite{Kauffmann_1993,Klypin_1999,Moore_1999a,Bode_2001}~\footnote{However, with better simulations and more careful characterisation of the observational selection effects, this discrepancy might been already resolved~\cite{Kim:2017iwr}.}. The second, known as the ``core-cusp problem,'' is the observed discrepancy between the inferred DM density profiles of low-mass galaxies and that predicted by CDM simulations~\cite{de_Blok_2010}. Simulations typically predict ``cuspy'' profiles (steep density profiles at the center) while observations suggest the existence of ``cores'' (softer density profiles at the center). 

While some of these apparent discrepancies between CDM models and observations could be attributed to astrophysical reasons (such as the effect of baryons), several new DM candidates have also been proposed to address them. WDM is the simplest departure from CDM, endowing the DM with a small velocity dispersion. WDM particle with a mass in $\mathrm{keV}$ range predicts the suppression of structures at small scales ($\sim 100\;\mathrm{kpc}$) without affecting the large scales ($\sim \mathrm{Mpc}$), thus explaining the missing satellites. Another model, called "self-interacting DM"~\cite{Spergel_2000}, adds a self-interaction cross-section to the DM. The resulting elastic scattering between the DM particles in the inner galactic regions redistributes energy, producing the effect of a core. Fuzzy DM (FDM) particles are ultralight bosons (mass $\sim 10^{-22}\;\mathrm{eV}$), with de Broglie wavelength larger than the inter-particle separation. The resulting wave-like behaviour leads to formation of solitonic cores at the center of haloes and density granules on scales smaller than $\sim \mathrm{kpc}$ are erased, while large scale structure is indistinguishable from CDM~\cite{Hu_2000}.

The  observation of Lyman-$\alpha$ forest --- absorption lines in the distant quasar spectra induced by neutral hydrogen along the line of sight --- provides the strongest lower limit  ($m_{\wdm}>3-5\; \mathrm{keV}$) on the WDM mass~\cite{Viel_2013a,markovic_viel_2014,Yeche_2017,Villasenor_2023}. Combining this with strong gravitational lensing~\cite{vegetti2023strong} and the abundance of Milky Way satellites has resulted in a joint constraint $m_{\wdm}>6.048\;\mathrm{keV}$~\cite{Enzi_2021}. {Different cosmological datasets puts upper bound $m_{\psi}\geq 10^{-22}\;\mathrm{eV}$ on the mass of FDM~\cite{Bozek_2015,Schive_2016, Corasaniti_2017, Ir_2017}. A stronger constraint, $m_{\psi}\geq(0.6-1)\times10^{-19}\;\mathrm{eV}$ is obtained from the survival of an old star cluster in an ultra faint dwarf galaxy Eridaus II \cite{Marsh_2019}}.  

Gravitational-wave (GW) observations offer new probes of DM (see, e.g.,~\cite{Bertone_2019}). The presence of a DM overdensity surrounding a black hole can have impact on the GWs emitted during the inspiral and merger with another compact object. The upcoming LISA observatory \cite{LISA} will be able to detect the effect of the DM on the GW signal, offering a powerful probe of the nature of particle DM~\cite{Bradley_2020,kadota2023,Bertone_2019}. {Ultralight bosons, such as axions, can affect the mass and spin of black holes by forming gravitationally bound states around them \cite{Brito_2015}. GW emission from such objects could be detected by various  GW detectors~\cite{Arvanitaki_2011,Arvanitaki_2015,Brito_2017, Baryakhtar_2017}. FDM can also be indirectly detected using pulsar timing arrays if the oscillation frequency falls within their detection band~\cite{Porayoko_2018}. 

Gravitational lensing of GWs offer yet another avenue to probe the nature of DM (see, e.g.~\cite{Dai_2018, Oguri_2020, Choi_2021, Gao_2022, Guo_2022, caliskan2023probing, Cao_2020, Cao_2022}, for some recent work). The next-generation (XG) ground-based GW detectors will detect millions of binary black-hole mergers (BBH) out to high redshifts ($z \sim 10-100$)~\cite{HallEvans3G}.  About $0.1 - 1\%$ of them will be strongly lensed by the galaxies and clusters hosted in these DM halos, producing multiple copies of the GW signals. The time delay between the lensed copies of these GW signals can be accurately measured. The exact fraction of lensed mergers and the  distribution of lensing time delay will depend on the mass distribution of lenses at various redshifts~\cite{Xu:2021bfn} as well as cosmological parameters~\cite{Jana_2023}. In this \emph{letter}, we propose a statistical probe to constrain the mass of the WDM particle using a catalogue of strongly lensed GW detections. If the DM is warm, this will hinder the formation of low-mass halos. This suppression in the the abundance of low-mass halos will result in a reduction in the number of lensed events with small time delays, as small time delays are mostly produced by low mass lenses.  

Our proposal is to look for the imprints of WDM on the number of lensed signals, as well as on the distribution of their time delays. This approach is closely connected to our earlier work~\cite{Jana_2023} on constraining the cosmological parameters from strongly lensed GW signals. Our method does not rely on the accurate knowledge of the source location of the individual signals or the properties of the corresponding lenses. Indeed, the number of lensed events as well as the time delay distribution will also depend on the distribution of source properties (e.g., mass and redshift distribution of BBHs \cite{haris2018,Li:2018prc,Mukherjee:2021qam}) as well as the lens properties (e.g., the mass function of the DM halos \cite{2020MNRAS.495.3727R} and the lens model \cite{2022arXiv220511499J,More:2021kpb}). If the distributions of the source and lens properties are known from other observations or theoretical models (e.g., from the observation of unlensed GW signals and cosmological simulations), then the mass of the WDM can be inferred from the observed number of lensed events and their time delay distribution. 

We forecast that BBH observations during $10\;\mathrm{yrs}$ of operation of XG  detectors~\cite{ETScienceCase,CosmicExplorer} will be to provide constraints ($m_{\wdm}^{-1} < 0.035 - 0.056 ~ \mathrm{keV}^{-1}$) are significantly better than the current constraints ($m_{\wdm}>3-5\; \mathrm{keV}$). We simply translate the constraints on $m_{\wdm}$ to mass of FDM particle ($m_{\psi}$). An optimistic assumption of merger rate will give us the constraint $m_{\psi} >7.3\times10^{-19} {\mathrm{eV}}$, which is almost three order of magnitude improvement over existing bounds.

\paragraph*{Warm dark matter:---}

\begin{figure}[tbh]
\includegraphics[width=0.7\columnwidth]{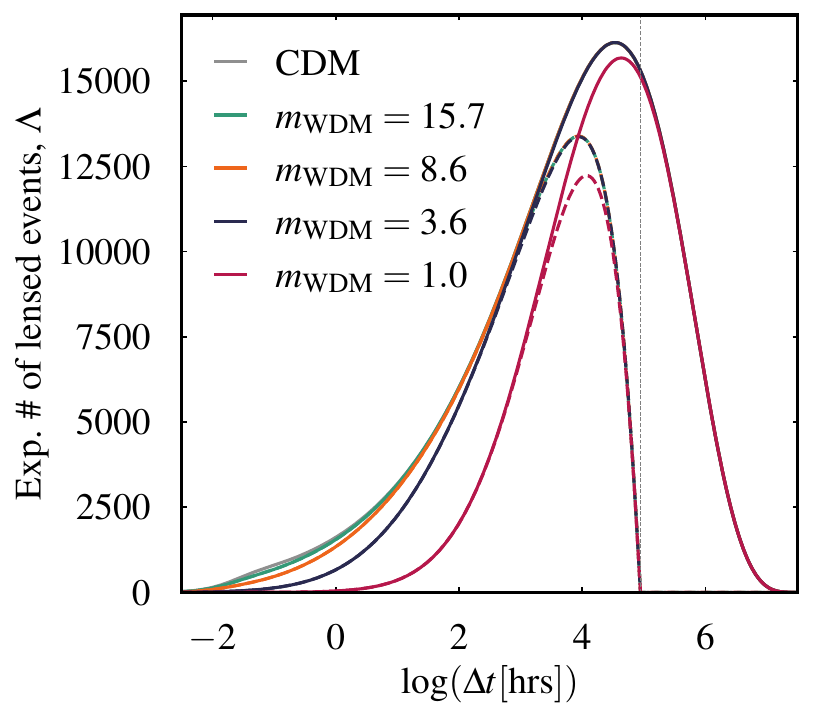}
\caption{Expected distributions of time delay between strongly lensed GW signals, corresponding to different values of $m_\wdm$. Note the suppression in number of lensed events compared to CDM, especially for lower time delays, which is the reflection of absence of the lower mass halos for smaller $m_\wdm$. The time delay distributions measurable from an observation period of $10\;\mathrm{yrs}$ are shown by dashed lines. The vertical line indicates the period of $10\;\mathrm{yrs}$.}
\label{fig:delta-t}
\end{figure}

Free streaming of WDM particles suppress primordial perturbations at scales smaller than the \emph{free streaming scale}. Fitting functions for modelling the WDM transfer function have been proposed in different studies~\cite{Bardeen_1986,Bode_2001,Viel_2005}. They give us a prescription to convert the power spectrum $P_\cdm(k)$ of linear perturbations in the CDM model to the same in the WDM model [$P_\wdm(k)$], through the transfer function. We use the transfer function given in \cite{Viel_2005},
\begin{equation}
\label{transfer-wdm}
T(k) =  \left[ \frac{P_\cdm(k)}{P_\wdm(k)} \right]^{1/2} = \left[1+(\alpha k)^{2\mu}\right]^{-5/\mu}, 
\end{equation} 
where $\mu=1.12$ and 
\begin{equation}
\alpha = 0.049 \left(\frac{m_\wdm}{\mathrm{keV}}\right)^{-1.11} \left(\frac{\Omega_\wdm}{0.25}\right)^{0.11} \left(\frac{h}{0.7}\right)^{1.22} ~ h^{-1} \mathrm{Mpc}
\end{equation} 
is called the \emph{effective free streaming scale}. Above, $\Omega_\wdm$ is the energy density in the form of WDM and $h$ is the Hubble constant in units of 100 km/s/Mpc. We can introduce another length scale, called \emph{half mode length scale} $\lambda_\mathrm{hm}$, where the WDM transfer function becomes half: $\lambda_{\mathrm{hm}}  = 2\pi\alpha \, (2^{{\mu}/{5}}-1 )^{-{1}/{2\mu}}$. This length scale introduces a \emph{half mode mass scale} $M_{\mathrm{hm}} = \frac{4\pi}{3}\bar{\rho} \,({\lambda_\mathrm{hm}}/{2} )^3$, where $\bar{\rho}$ is the averaged density of the halo. Abundances of DM haloes with mass below the $M_{\mathrm{hm}}$ will be suppressed compared to CDM, while the masses above $M_{\mathrm{hm}}$ are unaffected.

We use different halo mass functions (HMFs) that model the comoving number density $dn_\cdm/dM$ of CDM halos in different mass ranges. Given a HMF in CDM, the same in WDM model for a particular $m_\wdm$ is obtained using the fitting formula given in \cite{Schneider_2012} 
\begin{equation}\label{scale-to-wdm}
	\frac{dn_\wdm/dM}{dn_\cdm/dM} = \left(1+ \frac{M_{\mathrm{hm}}}{M}\right)^{-\beta},
\end{equation}
where $\beta = 1.16$.  The dependence on $m_\wdm$ comes through $M_\mathrm{hm}$. To obtain HMF in WDM model, we use \textsc{HMFcalc} package \cite{hmfcalc}. For our main analysis we consider the Behroozi~\cite{behroozi2013} model of the CDM HMF. In order to estimate the effect of using an incorrect CDM HMF model, we also consider the Jenkins~\cite{jenkins2001} model of the same implemented in the same package. 

These DM halos acts as gravitational lenses that deflect light as well as GWs. In this \emph{letter}, we are concerned about the strong lensing of GWs that produces multiple copies of the GW signals. We approximate these halo lenses as {singular isothermal spheres} (SISs)~\cite{Schneider_book}, parameterised by their dispersion velocity $\sigma$. We assume that the halos are spherically symmetric and virialised, with average density $\bar{\rho}$ and radius $R$. This allows us to compute the dispersion velocity of the SIS lens from the halo mass 
\begin{equation}
\sigma \simeq \sqrt{\frac{GM}{R}}, ~~~ M = \frac{4}{3}\pi R^3 \bar{\rho}. 
\label{eq:HMtosigma}
\end{equation}
Figure~\ref{fig:hmf-sigma} shows the the power spectrum, the HMF and the $\sigma$ distribution of lenses derived from the HMF, as predicted by the CDM model as well as the WDM model corresponding to different WDM masses. These can be use to compute the expected number of strongly lensed GW signals and the distribution of the time delay between the lensed copies of GW signals. 

Figure~\ref{fig:delta-t} shows the expected number of lensed events $\Lambda (m_\wdm)$ and the time delay distribution $p(\Delta t | m_\wdm)$ as a function of $m_\wdm$. We can see that $\Lambda$ decreases with decreasing $m_\wdm$. This is due to the fact that there will be a smaller number of low-mass halos for lower $m_\wdm$. The absence of lower mass halos is reflected in the time delay distribution as the reduction in the of lower time delays. Using these differences in the time delay distribution and total number of lensed events, we will be able to either measure the mass of the WDM particle, or put a lower bound on $m_\wdm$. In practice, we put an upper bound on $m_\wdm^{-1}$ since it has a convenient lower bound of zero.

\paragraph*{Expected constraints on warm DM:---}
\label{sec:results}

\begin{figure*}[tbh]
\includegraphics[width=2\columnwidth]{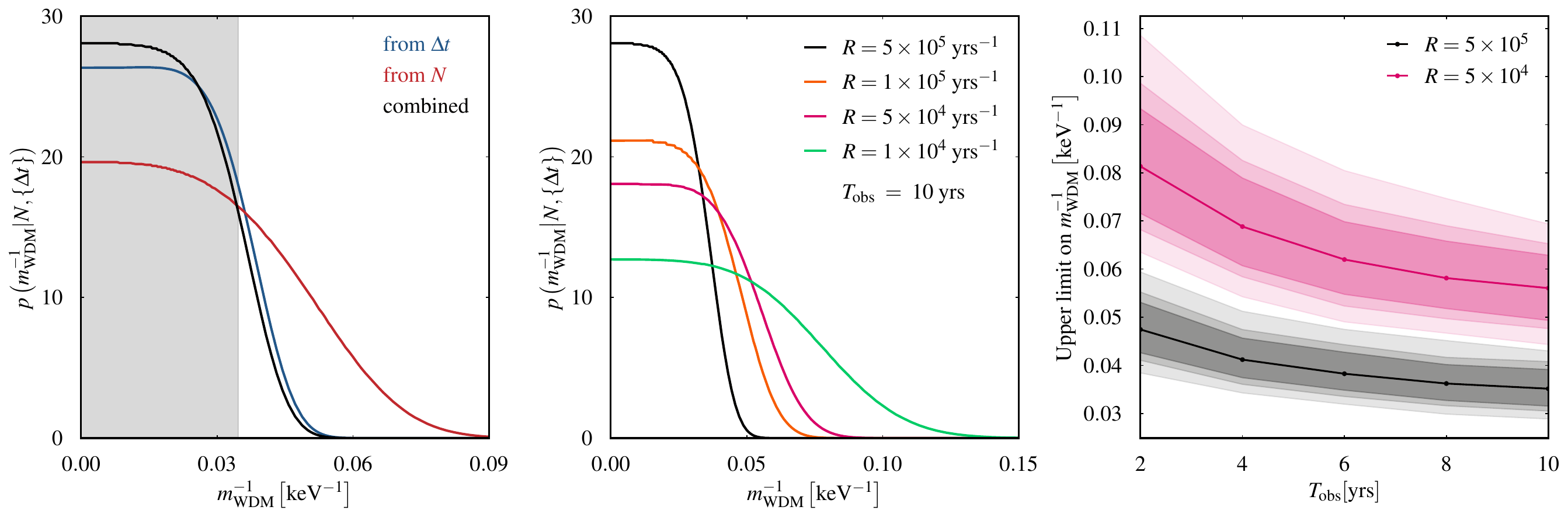}
\caption{\emph{Left panel:} Posterior distribution of $m_{\wdm}^{-1}$ computed from number of lensed events and time delay distribution separately, along with the combined posterior, using simulated observations generated assumming $\Lambda$CDM universe. Here we assumed a BBH detection rate of  $R=5\times10^5\;\mathrm{yr}^{-1}$ and $T_{\mathrm{obs}} = 10 \mathrm{yrs}$. Gray shaded region represents the $90\%$ quantile of the combined posterior, yielding an upper limit of  $m_\wdm^{-1} \leq 0.035$. \emph{Middle panel:} Combined constraints from GW lensing on $m_{\wdm}^{-1}$ assuming different detection rates with observation time period $T_{\mathrm{obs}}=10\;\mathrm{yrs}$.  \emph{Right panel:} 38\%, 50\% and 68\%  credible intervals (denoted by different shades) of the distributions of 90\% upper limit of $m_{\wdm}^{-1}$ obtained from $\sim1000$ recovery tests for different values of $R$ and $T_\mathrm{obs}$.}
\label{fig:mwdm-constraint}
\end{figure*}

We first ask the question: if the DM is actually cold (i.e., $m_\wdm^{-1} = 0$), how well can we constrain $m_\wdm^{-1}$ using future obervations of lensed GWs. In order to answer this, we simulate a population of BBH mergers with redshift distribution given by~\cite{dominik2013}. We assume a BBH detection rate $R = 5\times10^5\;\mathrm{yr}^{-1}$ and an observation period $T_{\mathrm{obs}} =10\;\mathrm{yrs}$. We neglect the selection effects in detection, as XG detectors are anticipated to detect all the BBH mergers out to large red shifts ($z \sim 10-100$). The mass distribution of the lenses at various redshifts are described by the CDM HMF model of \cite{behroozi2013}. Lenses as modelled using the SIS model, using Eq.~\eqref{eq:HMtosigma} for converting the halo mass to the velocity dispersion of the lens. This allows us to compute the the strong lensing optical depth for sources located at different redshifts, which can be convolved with the redshift distribution of the BBH mergers to to compute the expected number $\Lambda$ of lensed events.   We assume the following values of cosmological parameters: $\Omega_m = 0.316, H_0 = 67.3, \sigma_8 = 0.816~$\cite{Planck18}. 

To simulate one observing scenario with $N$ detections of lensed events, we draw one sample from a Poisson distribution with mean $\Lambda$. Further, we draw samples $\{\Delta t_i\}_{i=1}^N$ from $p(\Delta t~|~m_\wdm^{-1}\simeq0, T_{\mathrm{obs}})$. Using $N$ and $\{\Delta t_i\}_{i=1}^N$ we evaluate the likelihoods $p\left(N~|~m_\wdm^{-1},T_{\mathrm{obs}}\right)$ and $p\left(\{\Delta t_i\}~|~m_\wdm^{-1},T_{\mathrm{obs}}\right)$. We assume uniform priors on $m_{\wdm}^{-1}$, so final posterior is given by the product of two likelihoods. Figure~\ref{fig:mwdm-constraint} (left panel) shows the two likelihoods and the posteriror obtained from combining these two likelihoods.  The $90\%$ quantile of combined posterior is shown in the shaded region, yielding an upper limit $m_{\wdm}^{-1} < 0.035~\mathrm{keV}^{-1}$. 

The GW detection rate of XG detectors is uncertain as of now. For more conservative forecasts, we repeated the analysis with lower rates.  Figure~\ref{fig:mwdm-constraint} (middle panel) shows the combined posteriors for different detection rates. As expected, with reduced rates the posteriors are broader. The 90\% upper limits on $m_\wdm^{-1}$ in  are $0.047, 0.055, 0.082~\mathrm{keV}^{-1}$ for merger rates $10^5, 5 \times10^4, 10^4$, respectively. Constraints corresponding to different merger rates and observation periods are shown in Table~\ref{tab:mwdm-constraint-tobs}. Since the upper limits depend on the realization of mock samples that are subjected to Poisson fluctuations, we repeat the observing scenario $\sim1000$ times and compute the distribution of the 90\% limit of $m_\wdm^{-1}$. Figure~\ref{fig:mwdm-constraint} (right panel) shows the 38\%, 50\% and 68\% quantiles of the distributions of the 90\% upper limit of $m_{\wdm}^{-1}$ for different observation time periods and merger rates.

\begin{table}[h!]
	\centering
	\begin{tabular}{c | c c c c c} 
		\hline
		\hline 
		 \backslashbox[20mm]{$R$}{$T_{\mathrm{obs}}$}& $2\;\mathrm{yrs}$ & $4\;\mathrm{yrs}$ & $6\;\mathrm{yrs}$& $8\;\mathrm{yrs}$ &$10\;\mathrm{yrs}$\\ 
		 \hline
		 \hline 

		$5\times10^5\;\mathrm{yr}^{-1}$& 0.047 & 0.041 & 0.038& 0.036 &0.035\\ 
		\hline 
		$1\times10^5\;\mathrm{yr}^{-1}$& 0.068 & 0.059 & 0.053& 0.049 &0.047\\ 
		\hline
		$5\times10^4\;\mathrm{yr}^{-1}$& 0.081 & 0.069 & 0.062& 0.058 &0.055\\ 
		\hline
		\hline
	\end{tabular}
\caption{Expected 90\% credible upper limits on $m_{\wdm}^{-1}$ in $\mathrm{keV}^{-1}$ for different merger rates $R$ and observation time periods $T_\mathrm{obs}$.}
\label{tab:mwdm-constraint-tobs}
\end{table}

The FDM also predicts a cut-off in the HMF at small scales, through a mechanism that is dependent on the de Broglie wavelength rather than a free-streaming length. We translate the constraints on the $m_\wdm$ to the mass of the FDM particle $m_{\psi}$ (Fig.~\ref{fig:fdm-wdm-constraint}). For this, we simply equate the half mode length scale of both. We use the expression of half mode length scale of FDM as given in \cite{Schive_2016}  which considers a transfer function given in \cite{Hu_2000}. We calculate half mode length scale for WDM using the \textsc{HMFcalc} package \cite{hmfcalc} which uses the formula given in \cite{Schneider_2012} and a transfer function given in \cite{Viel_2005}. This is just a simple translation, in order to be more rigorous, one needs to use the HMF in FDM. Even though approximate, this gives us an idea of the prospective constraints on FDM using future observations of GW strong lensing.

\begin{figure}[tbh!]
\includegraphics[width=0.75\columnwidth]{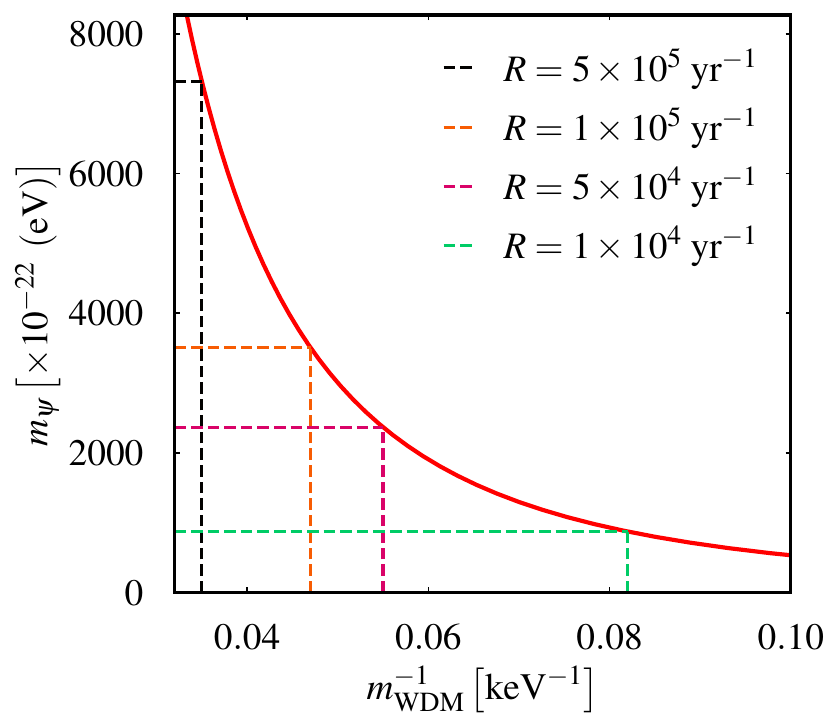}
\caption{The relation between $m_{\psi}$ and $m_{\mathrm{wdm}}^{-1}$  obtained by equating the half mode length scale of FDM and WDM. Dashed vertical line shows the constraints on $m_{\mathrm{wdm}}^{-1}$  and corresponding dashed horizontal lines show the translated constraints on $m_{\psi}$. Different dashed lines are for different detection rates $R$, assuming $T_\mathrm{obs} = 10\;\mathrm{yrs}$.} 
\label{fig:fdm-wdm-constraint}
\end{figure}

We also check whether we will be able to measure the mass of the DM particle when it is actually warm. To check this, we simulate an observing scenario using the HMF of the WDM model with mass  $m_\wdm=9~\mathrm{keV}$. Other details of the analysis are kept the same. We consider an optimistic ($R = 5 \times 10^5~\mathrm{yr}^{-1}$) and pessimistic ($R = 5 \times 10^4~\mathrm{yr}^{-1}$) detection rates. As seen in Fig.~\ref{fig:true-wdm}, the true value of $m_\wdm$ is recovered within 68\% credible interval. 
\begin{figure}[tbh]
\includegraphics[width=0.7\columnwidth]{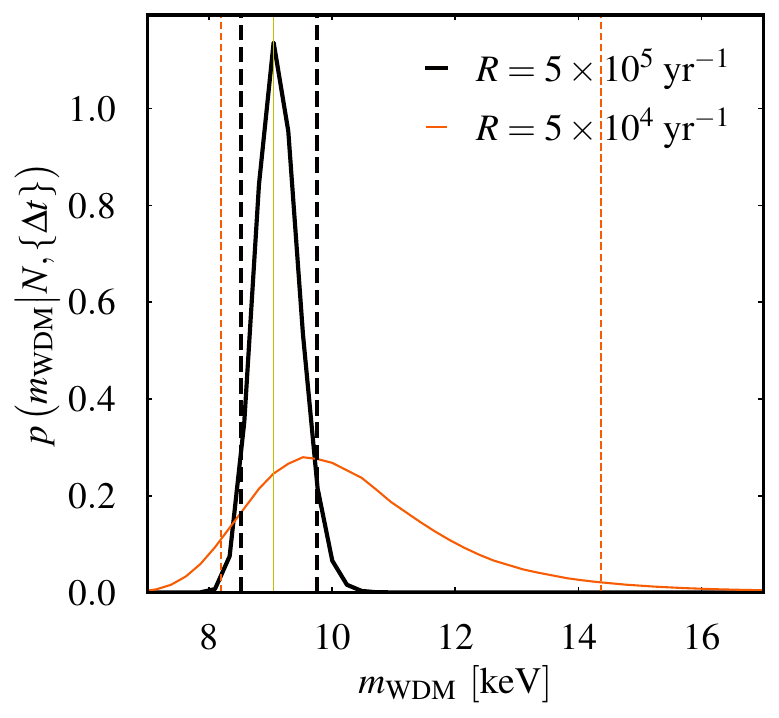}
\caption{Posterior distribution of $m_{\wdm}$ assuming that the true nature of the DM is described by the WDM model with $m_\wdm = 9~\mathrm{keV}$ (the yellow solid line in the middle). The plot shows the expected posteriors for $T_{\mathrm{obs}} = 10 \mathrm{yrs}$ assuming optimistic ($R = 5 \times 10^5~\mathrm{yr}^{-1}$) and pessimistic ($R = 5 \times 10^4~\mathrm{yr}^{-1}$) detection rates. Vertical dashed lines represent the 90\% credible region of the posteriors.}
\label{fig:true-wdm}
\end{figure}

\paragraph*{Systematic errors:---}
\label{systematics}

We have investigated various sources of systematic errors in deriving constraints on the nature of DM using future observations of strongly lensed GW signals. 

\begin{figure}[tbh]
\includegraphics[width=1\columnwidth]{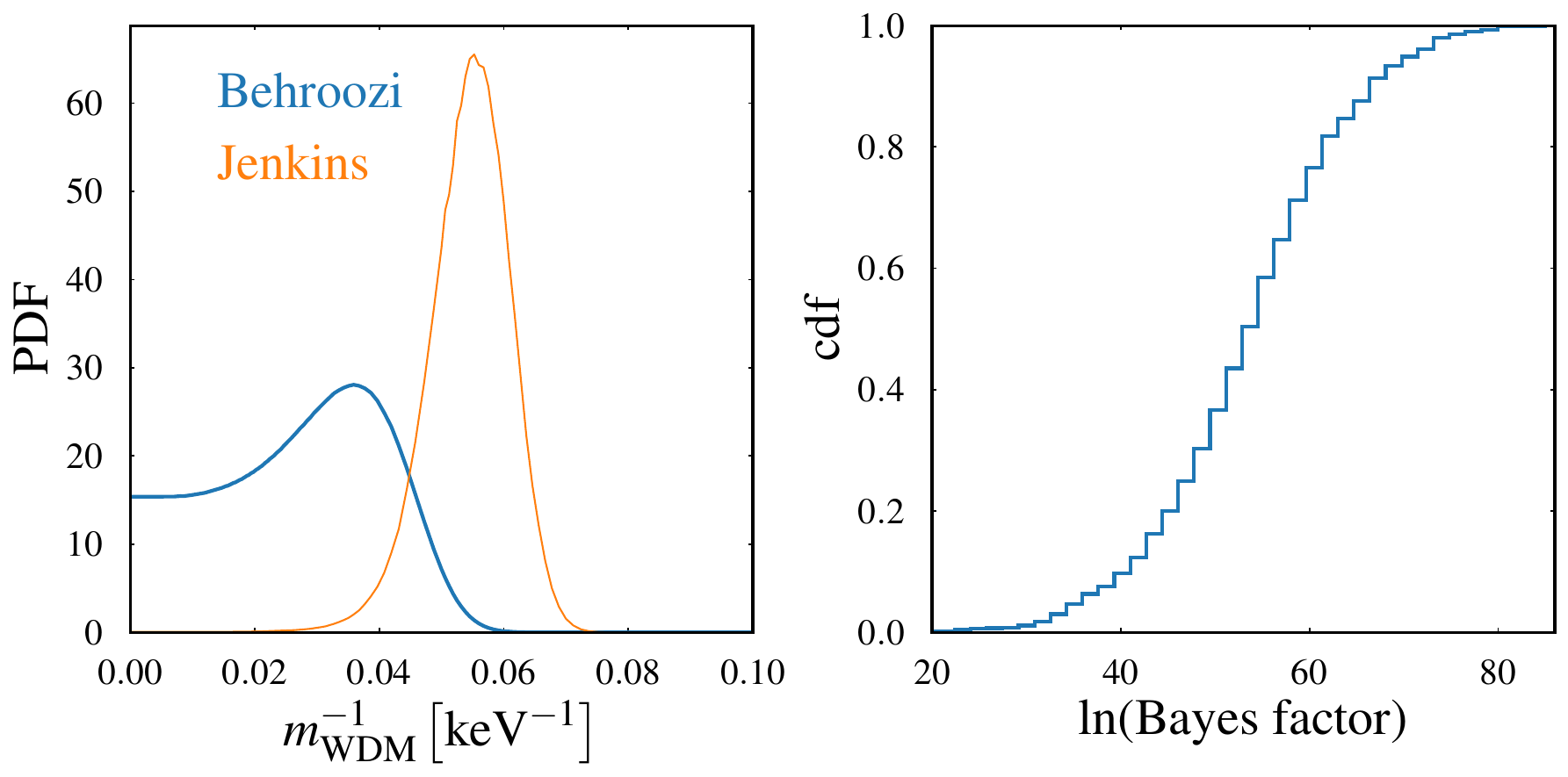}
\caption{\emph{Left panel:} Posterior distributions of $m_{\wdm}^{-1}$ computed from number of lensed events and time delay distribution. The lensed events are simulated assuming a CDM HMF ($m_\wdm^{-1} \simeq 0$) described by Behroozi model with  merger rate $R = 5 \times 10^5\;\mathrm{yr}^{-1}$ and $T_{\mathrm{obs}} = 10 ~\mathrm{yrs}$. The blue and orange lines show the posterior distribution of $m_\wdm^{-1}$ estimated using the Behroozi (``right'') and Jenkins (``wrong'') WDM HMF models, respectively. The true value of $m_\wdm^{-1} \simeq 0$ is not recovered when the wrong HMF model is employed in the inference. \emph{Right panel:} Cumulative distribution of the ratio of evidences between the Behroozi and Jenkins models. This shows ``right'' model (Behroozi) is consistently preferred over ``wrong'' model (Jenkins).}
\label{fig:bias-plot}
\end{figure}

One potential source of systematic error is the HMF that is used to model the mass and redshift distribution of lenses. To get a sense of the systematic errors, we simulate the population of lenses using one HMF model (Behroozi~\cite{behroozi2013}) and use another model (Jenkins~\cite{jenkins2001}) for our inference, both implemented in the \textsc{HMFcalc} package~\cite{hmfcalc}. As we see in left panel of Fig.~\ref{fig:bias-plot}, the true value ($m_\wdm^{-1} = 0$) is not recovered if we use the wrong HMF model in our inference. This underlines the need of accurate models of the distribution of lens properties.

Though the true value of $m_\wdm$ is not recovered in the parameter inference, we perform a Bayesian model selection study using these two HMF models to calculate the Bayes factor (ratio of Bayesian evidences) between them. We repeat this exercise over a large number of random realisations of the same observing scenario, and find that the Bayes factor overwhelmingly prefers the true HMF model (right panel of Fig.~\ref{fig:bias-plot}). Thus, if the correct HMF model is among the set of models that we consider for the parameter inference, we expect it to have the largest Bayesian evidence, thus will help us to evade systematic errors to a good extent. 

\begin{figure}[tbh]
\includegraphics[width=1.0\columnwidth]{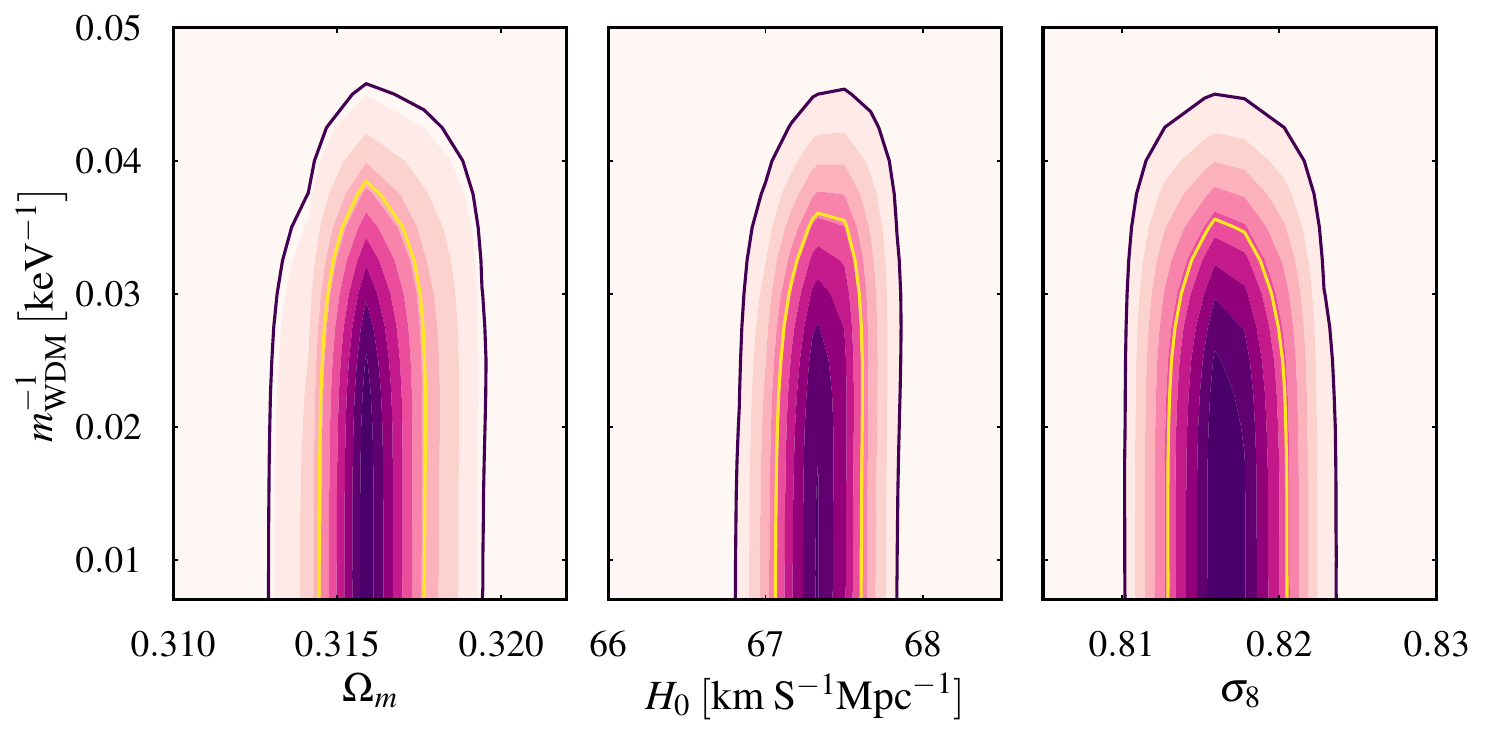}
\caption{The two-dimensional likelihoods of $m_\wdm^{-1}, \Omega_M, H_0$ and $\sigma_8$ that illustrate the correlation between these parameters. Each plot shows a 2-dimensional slice of the likelihood while keeping all other parameters to their true value. The colors represent the value of likelihood and the contour lines shows 68\% and 95\% credible levels. It can be seen that there is no correlation between $m_\wdm^{-1}$ and the cosmological parameters.}
\label{fig:cosmo-correlation}
\end{figure}

We also check whether the change in the expected number of strongly lensed events and their time delay distribution due to differences in the DM models could be mimicked by changes in cosmological parameters as studied in~\cite{Jana_2023,Jana_2024}. If such a degeneracy exists, the properties of the DM particle will be indistinguishable from cosmological parameters. To study this, we generate a sample of lensed BBH events assuming the standard value of the cosmological parameters that we used in the paper along with the CDM HMF. We then compute the likelihood by simultaneously varying $m_\wdm$ as well as the cosmological parameters $H_0, \Omega_M$ and $\sigma_8$. Figure \ref{fig:cosmo-correlation} shows various 2-dimensional slices of the likelihood that illustrate the lack of correlation between $m_\wdm$ and cosmological parameters. This suggests that marginalising over the cosmological parameters will not change the posterior of $m_\wdm^{-1}$ significantly. We leave a joint analysis including $m_\wdm^{-1}$ and all other cosmological parameters for future work.

The effect of WDM that we consider is the suppression in the number density of low mass halos. Besides this effect, WDM predicts the existence of finite central cores in DM halos \cite{tremaine1979,Bode_2001}, as opposed to the cuspy profile expected in CDM. The presence of a central core reduces the Einstein angle of the lens, resulting in smaller time delays as compared to halos without core. This effect can be populate the low-time delay part of the time-delay distribution, potentially mimicking CDM halos. However, the presence of cores will also reduce the lensing optical depth. Our preliminary investigations suggest that the resulting reduction in the number of lensed events should make them distinguishable from the CDM halos. This is being investigated in a follow-up work. 

\paragraph*{Summary and Outlook:---}
\label{sec:conclusion}

The large number of strongly lensed GW signals detectable by the XG detectors will enable new probes of cosmology. In this \emph{letter}, we showed how the expected number of strongly lensed BBH mergers and the distribution of the time delay between lensed images can be used to constrain the nature of particle DM. In particular, the suppression of the low-mass DM halos predicted by the WDM/FDM models will result in a reduction in the number of strongly lensed GW signals with small lensing time delays. This will allow us to constrain the mass of the WDM/FDM particle. The expected constraints are significantly better than the current bounds. In the time delay distribution, there is no degeneracy between the nature of DM particle and the cosmological parameters. This should allow us to constrain both the nature of DM and cosmological parameters simultaneously from future observations. 

Because of their inherent simplicity, GWs are unaffected by extinction, and selection effects in GW searches are well modelled. This makes GW strong lensing a cleaner probe than, e.g., optical lensing. However, a number of obstacles need to be addressed before this technique can be applied to constrain the nature of DM. The properties of astrophysical sources and lenses determine both the number of lensed events and the distribution of their time delays. Some of the relevant parameters, such as the redshift distribution of BBH mergers, can be inferred from the large number of unlensed GW signals as well as the stochastic GW background. For other parameters, such as the distribution of the properties of the lens parameters, will need to rely on models obtained from large-scale cosmological simulations and galaxy surveys. 
		
In this \emph{letter}, we have assumed that the lenses are modelled by simple SISs, whose parameters are obtained from the HMF using a simple prescription. We also neglected the halo sub structure and the effect of baryons. Future analysis needs to model the lenses more accurately, which are working on improving~\cite{Jana_InPrep}.  We neglected selection function of the detectors assuming that XG detectors will observe all the BBHs out to large redshifts. In order to forecast the expected constraints from the upcoming upgrades of the current generation detectors~\cite{Post_05,Voyager_whitepaper_2023,Voyager_2020}, we will need to take into account the selection effects also, which is being studied in another ongoing work~\cite{Maity_InPrep}.  There are other potential sources of systematics: The false positives produced by the methods used to identify strongly lensed signals can contaminate our observational sample and can bias the parameter inference. In \cite{Jana_2024} we demonstrated how we can model the contamination thus evading systematic biases in the estimation of cosmological parameters from strongly lensed GW signals. A similar approach can be adopted here as well. 

While we have focussed on BBH signals, strongly lensed binary neutron star mergers, which are expected to be equally abundant~\cite{Baibhav_2019,magare2023,Smith_2023}, could also be used to probe cosmological parameters and the nature of DM. In contrast to BBH mergers which can be observed out to very large redshifts ($z \sim 10$), neutron star mergers can be observed out to smaller redshifts ($z \sim 1$) even using XG detectors. However, some of them would have an electromagnetic counterpart which will also be lensed. Such observations could allow us to probe the detailed profile of the lensing galaxy, potentially enabling to put tighter constraints on the nature of DM. This is being explored in an ongoing work~\cite{Maity_InPrep2}.

\paragraph*{Acknowledgements:} \label{sec:ack}
We are grateful to Jay Wadekar and Otto Akseli Hannuksela for useful discussions and comments. We are grateful to Steven Murray for his valuable assistance with the HMFcalc package. We also thank the members of the ICTS Astrophysics \& Relativity group for useful discussions. Our research was supported by the Department of Atomic Energy, Government of India, under Project No. RTI4001. SJK’s work was supported by a grant from the Simons Foundation (677895,  R.G.).  PA’s research was supported by the Canadian Institute for Advanced Research through the CIFAR Azrieli Global Scholars program.   {TV acknowledges support from NSF grants 2012086 and 2309360, the Alfred P. Sloan Foundation through grant number FG-2023-20470, the BSF through award number 2022136, and the Hellman Family Faculty Fellowship.} All computations were performed with the aid of the Alice computing cluster at the International Centre for Theoretical Sciences, Tata Institute of Fundamental Research.

	\bibliography{references}
	
	\clearpage
	\onecolumngrid
	\begin{center}
		\textbf{\large Supplemental Material:\\ Probing the nature of dark matter using strongly lensed gravitational waves from binary black holes}
	\end{center}
	\twocolumngrid
	\setcounter{equation}{0}
	\setcounter{figure}{0}
	\setcounter{table}{0}
	\setcounter{page}{1}
	
	\section{Bayesian inference}
	We assume that we have confidently detected $N$ strongly lensed BBH events within an observation period of $\tobs$, each producing two observable copies (lensed images). We aim to compute the posterior distribution of the mass $m_\wdm$ of the WDM particle, using the time delays $\left\{\Delta  t_i\right\}_{i=1}^N$ from the $N$ detected lensed events. 
	
	We can write the likelihood $p\left(N,\left\{\Delta t_i\right\}~\big|~\mwdm^{-1},\tobs\right)$ as a product of two likelihoods as $N$ and $\left\{\Delta t_i\right\}$ are uncorrelated. The likelihood of observing $N$ events can be described as a Poisson distribution
	\begin{equation}
		p\left(N~\big|~\mwdm^{-1},\tobs\right) = \frac{\Lambda\left(\mwdm^{-1}\right)^N e^{-\Lambda\left(\mwdm^{-1}\right)}}{N!},
	\end{equation}
	where $\Lambda\left(\mwdm^{-1}\right)$ is the expected number of lensed events for a given value of $\mwdm^{-1}$.
	The likelihood of observing a set of time delays $\left\{\Delta t_i\right\}$ can be written as 
	\begin{equation}
		p\left(\left\{\Delta t_i\right\}~\big|~\mwdm^{-1},\tobs\right) = \prod_{i=1}^{N} p\left(\Delta t_i~\big|~\mwdm^{-1},\tobs\right). 
	\end{equation}
	Here we assume each BBH merger to be an independent event and that the time delays are measured accurately and precisely. $p\left(\Delta t_i~\big|~\mwdm^{-1},\tobs\right)$ is the ``model" time delay distribution calculated at measured time delay $\Delta t_i$. The model time delay distribution is calculated from the expected time delay distribution $p\left(\left\{\Delta t_i\right\}~\big|~\mwdm^{-1}\right)$, considering the fact that time delays greater than the observation time period $\tobs$ cannot be observed, see \cite{Jana_2023,Jana_2024} for more details. Evaluation of the likelihood, $p\left(N,\left\{\Delta t_i\right\}~\big|~\mwdm^{-1},\tobs\right)$ requires the expected number of lensed events, $\Lambda$ and expected time delay distribution, $p\left(\left\{\Delta t_i\right\}~\big|~\mwdm^{-1}\right)$ for different values of $\mwdm^{-1}$. These quantities depend on the values of cosmological parameters and the distribution of the source and lens properties \cite{Jana_2023,Jana_2024}. Here we fix the cosmological parameters to the Planck values, $\vec{\Omega}_P=\{\Omega_m = 0.316, H_0 = 67.3, \sigma_8 = 0.816\}$\cite{Planck18}. 
	We also consider the distribution of source and lens properties is obtained by other observations, theoretical studies and large scale cosmological simulations.
	
	\section{Calculation of expected number of lensed events and time delay distribution }
	
	We calculate the expected number of lensed events $\Lambda\left(\mwdm^{-1},\tobs\right)$ and the expected time delay distribution $p\left(\Delta t~\big|~\mwdm^{-1}\right)$ using equations 7 and 17 of \cite{Jana_2024}. The expected number of lensed events and their time delay distribution depends on $\mwdm^{-1}$ through the halo mass function.
	All the assumptions and inputs (source and lens distributions), used to calculate the total number of lensed events and their time delay distribution are same as outlined in \cite{Jana_2023}. We assume that the source redshift distribution, $p_b(z_s)$ will be known with adequate precision from the observation form unlensed events which will dominate the future dataset. For illustration, we use the source redshift distribution provided by the source population model \cite{dominik2013}, assuming standard cosmology, $\vec{\Omega_p}$. We model the distribution of lenses using the halo mass function, which gives the distribution, $p\left(\sigma,\zl\right)$ of $\sigma$ and $\zl$. We consider a mass range of $10^8$ to $10^{15}\; M_{\odot}$ for the lenses. For further details on the calculations, refer to \cite{Jana_2023,Jana_2024} and for insights into how the time delay distributions and the total number of lensed events are affected by changes in the assumptions about source population and lens distributions, please refer to \cite{Jana_2024}. We consider halo mass function model described in \cite{behroozi2013}. To check the bias incurred in the inference on the $\mwdm^{-1}$ by using wrong model of halo mass function we also consider another model of halo mass function described by \cite{jenkins2001}. 
	
\end{document}